\documentclass[aps,prl,showpacs,twocolumn,groupedaddress,superscriptaddress,
nobibnotes]{revtex4-2}

\usepackage{blindtext}

\usepackage[colorlinks,
          linkcolor=blue,
            citecolor=blue,
            urlcolor=blue
           ]{hyperref}
\usepackage{amsmath,amsthm,amssymb}
\usepackage{array}
\usepackage{subfigure}
\usepackage{graphicx}
\usepackage{fancyhdr}
\usepackage{color}
\usepackage{extarrows}
\usepackage{enumerate}
\usepackage{epstopdf}
\usepackage{bm}
\usepackage{natbib}
\usepackage{comment}
\usepackage{float}
\usepackage{ulem}

\usepackage{lipsum}

\newcommand {\B}{\textcolor {blue}}

\def\ncto{Na$_2$Co$_2$TeO$_6$\xspace}

\begin{document}

%\title{Emergent quantum disordered phase in Na$_2$Co$_2$TeO$_6$ under intermediate magnetic field along $c$ axis}
%\title{Field-induced Intermediate Quantum Disordered Phase in Kitaev Magnet Na$_2$Co$_2$TeO$_6$}
\title{Dominant Kitaev Interaction and Field-induced Quantum Disordered Phase\\ in the Cobaltate Na$_2$Co$_2$TeO$_6$}

\author{Xu-Guang~Zhou}
\thanks{These authors contributed equally to this work.}
\affiliation{Institute for Solid State Physics, University of Tokyo, Kashiwa, Chiba 277-8581, Japan}
\affiliation{School of Physics, Southeast University, Nanjing, Jiangsu, 211189, China}
\affiliation{Anhui Key Laboratory of Low-Energy Quantum Materials and Devices, High Magnetic Field Laboratory, Hefei Institutes
of Physical Science, Chinese Academy of Sciences, Hefei 230031, China}

\author{Han~Li}
\thanks{These authors contributed equally to this work.}
\affiliation{School of Physical Science and Engineering, Beijing Jiaotong University, 
Beijing 100044, China}

\author{Chaebin~Kim}
\thanks{These authors contributed equally to this work.}
\affiliation{Center for Quantum Materials, Seoul National University, Seoul 08826, Republic of Korea}
\affiliation{School of Physics, Georgia Institute of Technology, Atlanta, Georgia 30332, USA}

\author{Akira~Matsuo}
\affiliation{Institute for Solid State Physics, University of Tokyo, Kashiwa, Chiba 277-8581, 
Japan}

\author{Kavita~Mehlawat}
\affiliation{Institute for Solid State Physics, University of Tokyo, Kashiwa, Chiba 277-8581, 
Japan}

\author{Kazuki~Matsui}
\affiliation{Institute for Solid State Physics, University of Tokyo, Kashiwa, Chiba 277-8581, 
Japan}

\author{Zhuo~Yang}
\affiliation{Institute for Solid State Physics, University of Tokyo, Kashiwa, Chiba 277-8581, 
Japan}

\author{Atsuhiko~Miyata}
\affiliation{Institute for Solid State Physics, University of Tokyo, Kashiwa, Chiba 277-8581, 
Japan}

\author{Gang~Su}
\affiliation{Institute of Theoretical Physics, Chinese Academy of Sciences, Beijing 100190, China}
\affiliation{Kavli Institute for Theoretical Sciences, University of Chinese Academy of Sciences, Beijing 100190, China}

\author{Koichi~Kindo}
\affiliation{Institute for Solid State Physics, University of Tokyo, Kashiwa, Chiba 277-8581, Japan}

\author{Je-Geun~Park}
%\email{ymatsuda@issp.u-tokyo.ac.jp}
\affiliation{Center for Quantum Materials, Seoul National University, Seoul 08826, Republic of Korea}
\affiliation{Department of Physics and Astronomy, Institute of Applied Physics, Seoul National University, Seoul 08826, Republic of Korea}

\author{Yoshimitsu~Kohama}
\email{ykohama@g.ecc.u-tokyo.ac.jp}
\affiliation{Institute for Solid State Physics, University of Tokyo, Kashiwa, Chiba 277-8581, Japan}

\author{Wei~Li}
\email{w.li@itp.ac.cn}
\affiliation{Institute of Theoretical Physics, Chinese Academy of Sciences, Beijing 100190, China}

\author{Yasuhiro~H.~Matsuda}
\email{ymatsuda@issp.u-tokyo.ac.jp}
\affiliation{Institute for Solid State Physics, University of Tokyo, Kashiwa, Chiba 277-8581, 
Japan}

\begin{abstract}
The identification of quantum spin liquid phases in Kitaev candidate materials remains a major experimental challenge. Since most Kitaev candidates develop antiferromagnetic (AFM) order at low temperatures, currently there are great interest on the field-induced magnetic disordered phase in these compounds, that are distinct from (partially) polarized states. Recently, a cobaltate Na$_2$Co$_2$TeO$_6$ has emerged as a promising Kitaev candidate with high-spin $t^{5}_{2g}e^2_g$ configuration and spin-orbit entangled $J_{\rm eff} = 1/2$ honeycomb lattice system. There are intensive studies on field-induced magnetic states and phase transitions under in-plane magnetic fields. In this study, we propose an intermediate disordered phase induced by an out-of-plane field along the $c$-axis, through high-field magnetization and magnetocaloric effect measurements. 
To explain the high-field behavior of Na$_2$Co$_2$TeO$_6$, we develop an effective $K$-$J$-$\Gamma$-$\Gamma^{\prime}$ spin model featuring a dominant AFM Kitaev interaction. This framework uncovers an intermediate quantum spin liquid phase, establishing the material as a unique platform for exploring Kitaev physics and field-induced quantum-disordered states.
\end{abstract}

\maketitle

The spin-1/2 Kitaev honeycomb model, characterized by its bond-dependent nearest-neighbor interactions, is remarkable for hosting an exact quantum spin liquid (QSL) ground state~\cite{Kitaev2006}. Such quantum disordered states, particularly that under magnetic fields, feature long-range entanglement and fractional excitation, and constitute major platform for topological quantum computation~\cite{Nayak2008,Lahtinen2017}. Over the past decades, great experimental efforts have been devoted to searching for ideal realization of the Kitaev 
model in honeycomb layered compounds, particularly those with 4$d$ and 5$d$ transition-metal ions~\cite{Jackeli2009,Chaloupka2010,Takagi2019}. These studies also highlight that the Heisenberg interaction ($J$) and off-diagonal terms ($\Gamma$, $\Gamma^{\prime}$) must be included to understand the magnetic ordering in these compounds~\cite{Kubota2015,Johnson2015,Kasahara2018}. Nevertheless,  experimental signatures of QSL have been reported in Kitaev candidate materials under in-plane magnetic fields~\cite{Kasahara2018,Zheng2017,Sears2017,Winter2018,Jiang2019,JW2019QMats,Baek2017,Ran2017,Leahy2017,Ponomaryov2020,Kasahara2018Unusual,Yokoi2021Science,Banerjee2018,Banerjee2016,Banerjee2017,Do2017}, while the observed effects are often subtle and highly sensitive to external field~\cite{Zhao2022neutron}, requiring more stringent experimental verification.

\begin{figure}[t!]
\begin{center}
\includegraphics[width = 1.0\linewidth]{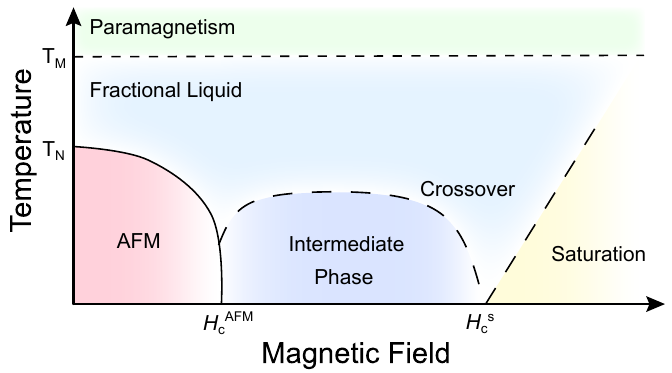}
\caption{Schematic field-temperature phase diagram: Under small fields, the ground state is an AFM phase due to the additional non-Kitaev terms. Above the critical field suppressing the AFM state, the system enter an intermediate quantum disordered phase, caused by the dominant Kitaev interaction. As temperature increasing, the quantum intermediate phase cross over to the fractional liquid regime~\cite{Nasu2014}, which finally enters the paramagnetic phase at high temperature. Under high field and at low temperature, there exists a saturation regime. The solid boundary denote the AFM transition, while the dashed lines label the crossovers between different regimes.}
\label{Idea}
\end{center}
\end{figure}

Moreover, the in-plane field-induced QSL phase has not been well supported by theoretical studies based on the $K$-$\Gamma$-$\Gamma^{\prime}$-$J$ spin model~\cite{Kaib2019, Jiang2019, Jiucai2019a, Gordon2019, Han2021}. Theoretical discussions on the in-plane field-induced QSL in candidate materials have so far been limited to the sixfold symmetry observed in specific heat, which may be linked to the Majorana gap in the Kitaev system~\cite{Tanaka2022thermodynamic,Fang2025,Hwang2022}. In contrast, various calculations predict the existence of a robust intermediate QSL phase~\cite{Yadav2016,Kaib2019,Jiang2019,Chern2020,Wang2019,Gordon2019, Jiucai2019a, Han2021,Zhang2022} under out-of-plane magnetic fields beyond the perturbative regime. The schematic field-temperature ($H$-$T$) phase diagram is presented in Fig.~\ref{Idea}.

Under out-of-plane fields, for candidates with dominant Kitaev interactions, two types of robust intermediate phases are predicted in the $H$-$T$ phase diagram: the finite-$T$ fractional spin liquid phase~\cite{Motome2020,Nasu2014,Kubota2015,Kavita2017} and the field-induced QSL phase~\cite{Yadav2016, Kaib2019,Jiang2019,Chern2020,Wang2019,Han2021,Gordon2019,Zhang2022}. While theoretical studies have established a robust framework for identifying out-of-plane intermediate QSL states, a systematic exploration of the complete phase diagram is still limited~\cite{Zhou2023possible,Yang2022,Modic2021}, which is crucial for addressing the fundamental questions on the field-induced intermediate phase. However, achieving accurate experimental results requires several critical conditions: the presence of dominant Kitaev interactions (challenging to confirm through experiments like the inelastic neutron scattering), the mechanical robustness which is necessary to against the stress caused by magnetic torque~\cite{Cao2016, Zhou2023possible, Yang2022}, and sufficiently high measurement precision within the ultra-high critical magnetic field range~\cite{Yang2023,Zhang2023,Zhou2023possible}.

Na$_2$Co$_2$TeO$_6$ is a very promising Co-based Kitaev candidate material that have raised great research interest~\cite{lin2021nc,pilch2023field, Kim2022,Yang2022,Wilhelm2023,Lukas2024,Jin2025}. The 3$d^7$ Co$^{2+}$ ions, with a high-spin $t^{5}_{2g}e^2_g$ configuration and a spin-orbit-entangled $J_{\rm eff}$ = 1/2 state, form a honeycomb lattice believed to possess Kitaev coupling~\cite{lin2021nc,Sano2018,Liu2020PRL,Liu2018PRB,Kim2022}. Zero-field specific heat experiments find two specific heat peaks in this material, which are located at T$_{\rm N}$ ($\sim$28~K) and T$_{\rm M}$ ($\sim$100~K)~\cite{Yang2022,Yao2022PRL}, respectively. The former corresponds to a magnetic phase transitions, while the latter corresponds to a crossover (see Fig.~\ref{Idea}). Moreover, inelastic neutron diffraction and terahertz spectroscopy experiments have observed a continuum of magnetic excitation~\cite{pilch2023field,Songvilay2020}, providing dynamical evidence for the presence of Kitaev interactions. Despite ongoing debates about the strength of Kitaev coupling in Na$_2$Co$_2$TeO$_6$~\cite{pilch2023field,xiang2023disorder,Kim2022,lin2021nc,Samarakoon2021PRB,Songvilay2020,Yao2022PRL}, the intriguing quantum magnetic behaviors in this material, along with its 
excellent mechanical robustness, makes it a highly promising platform for exploring the out-of-plane field-induced spin states and transitions.

In this study, by performing magnetization ($M$) measurement up to 100~T and magnetocaloric 
effect (MCE) measurement up to 55~T under the $c$-axis external field, we propose a field-induced quantum disordered phase between two critical fields at $H_c^{\rm AFM}\sim$37~T and $H_c^{\rm S}\sim$82~T in Na$_2$Co$_2$TeO$_6$.
We also tilt the angle $\theta$ of the magnetic field from the $c$-axis, and present the low-temperature $\theta$-$H$ phase diagram of Na$_2$Co$_2$TeO$_6$, which is found in stark similarity to that of $\alpha$-RuCl$_3$ reported in Ref.~\cite{Zhou2023possible}.
To understand the experimental observations, we propose an effective microscopic spin model with large antiferromagnetic (AFM) $K>0$ interaction. Remarkably, the model not only reproduces our experimental data but also indicates \ncto may host a intermediate-field QSL states and have connections with the results of a pure AFM Kitaev model under out-of-plane fields. Most importantly, these findings establish \ncto as a prime candidate for exploring QSL physics in Kitaev magnets, calling for further high-field out-of-plane studies to fully characterize this exotic phase.

\begin{figure}[b!]
\begin{center}
\includegraphics[width = 1.0\linewidth]{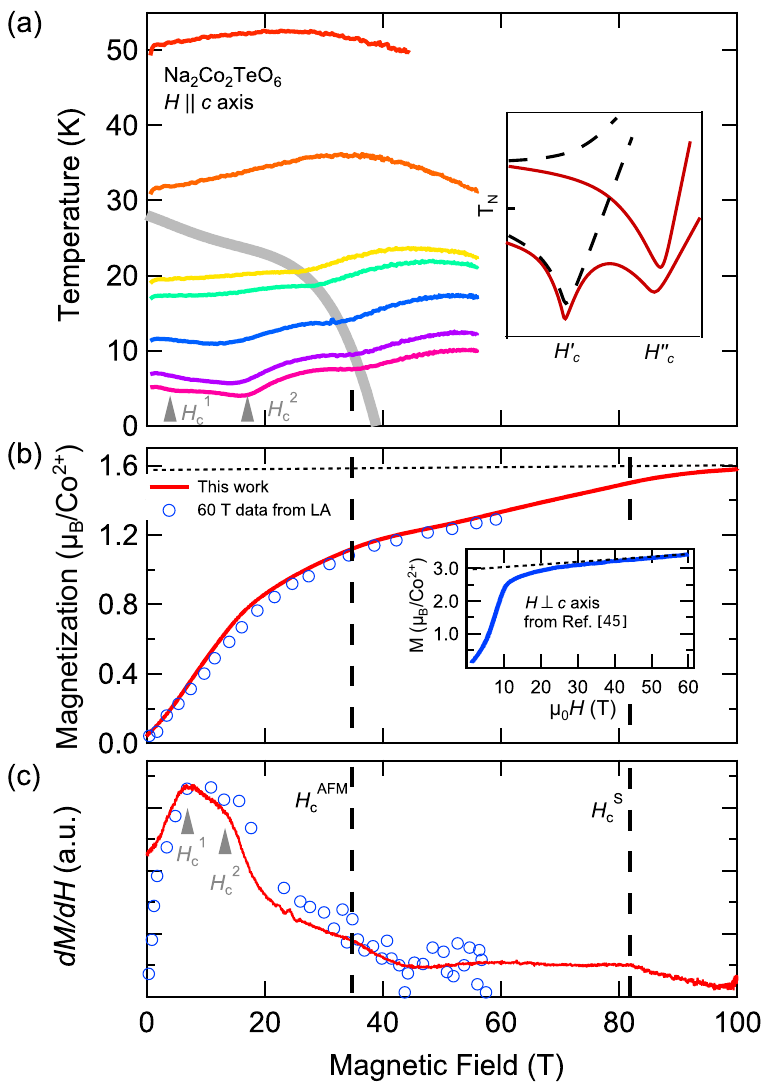}
\caption{
(a) The MCE measurements under non-destructive magnetic fields. The field was applied along the $c$ axis (tilting angle $\theta$ = 0$^{\circ}$); (b) Magnetization and (c) d$M$/d$H$ data measured along the out-of-plane field ($H // c$-axis) up to 100~T. The critical field, $H_c^1$ and $H_c^2$ are marked with gray arrows. Critical field $H_c^{\rm AFM}$ naturally defines a phase boundary (indicated by the gray curve) in (a), and also corresponds to magnetization data denoted by the black dashed line in (b) and (c). The critical field $H_c^{\rm S}$ is also marked by a black dashed line in (b) and (c). The blue circles represent the magnetization curve up to 60~T adopted from Ref.~\cite{Zhang2023}. The inset of (a) is a schematic plot to explain the different MCE features of two scenarios: a single transition from AFM to saturation regime (black dashed curves), and two transitions with an intermediate phase (red solid curves). The inset of (b) shows the magnetization results under in-plane magnetic field also adopted from Ref.~\cite{Zhang2023}, where the nonzero slope of the dashed fitting line reflects the van Vleck paramagnetism.}
\label{Temp}
\end{center}
\end{figure}

High-quality single crystal of Na$_2$Co$_2$TeO$_6$ were grown by a flux method~\footnote{A platinum crucible is used in this experiment. The temperature and other synthesis sequence adhered to the same workflow as detailed in Ref~\cite{Kim2022}, and we obtained dark-red thin single crystals}. The external magnetic fields up to 100~T~\cite{Miura2003research} and 55~T are generated by vertical-type single-turn coil and non-destructive field generators, respectively. The magnetization processes under out-of-plane fields and those at various rotated angles were measured using a 1.6 mm diameter pick-up coil consisting of two small coils compensating for each other~\cite{Takeyama2012,Matsuda2013,Zhou2020,Zhou2023possible}. The field directions of magnetization measurements are controlled in a similar manner as described in Ref.~\cite{Zhou2023possible}. Weak transitions below 50 T were confirmed by non-destructive magnetization measurements. All magnetization measurements were conducted at 4.2~K. In the MCE measurements, the field dependence of the sample temperature was measured in pulsed magnetic field using a AuGe thin film thermometer~\cite{kihara2013adiabatic}. We also perform the density matrix renormalization group (DMRG) method~\cite{PhysRevLett.69.2863} to fit with the experimental data. More details of the method and results could be found in Supplemental Material \B{B}~\footnote{See Supplemental Material [url] for additional
 experimental and calculated details, which includes Refs.~\cite{Kim2022,
 LinG2022,Samarakoon2021PRB,Songvilay2020,Yao2022PRL,Han2021,Bera2017,
 Yang2022,Yao2020PRB,Chaloupka2015,li2025kitaevderivedgaplessspinliquid,Lefran2016zigzag,Yu2023}.\label{SMREF}}.

In Fig.~\ref{Temp} (a), we present the measured MCE for $H\parallel c$ up to 55~T.
The dips in the isentropic $T$-$H$ curves sensitively signal the field-induced 
phase transitions and can be used to map the temperature-field phase 
boundaries~\cite{kohama2019possible, Nomura2020prb, Wang2019prl}. As shown in 
Fig.~\ref{Temp} (a), the isentropic curve starting from 5~K at zero field exhibits three 
local minima at $H_c^1 \simeq 4$~T, $H_c^2 \simeq 16$~T, and $H_c^{\mathrm{AFM}} \simeq 37$~T. 
Prior work 
associates $H_c^1$ and $H_c^2$ with either a magnetic 
plateau~\cite{Zhang2024outofplane} or weak unequal spin 
canting~\cite{arneth2025competing, Takeda2022, Zhang2023, Xiao2021magnetic}, which may
be related to the interlayer interaction~\cite{Samarakoon2021PRB, Chen2021PRB}. Here, 
we focus on the AFM transition (gray 
curve in Fig.~\ref{Temp}(a)), starting from the zero-field N\'eel temperature 
$T_\mathrm{N}$. The critical field $H_c^{\mathrm{AFM}}$ shifts towards higher values as
temperature lowers, smoothly connecting to the zero-temperature quantum phase 
transition and reflecting the field-induced suppression of the AFM order.

Previous work~\cite{Zhang2024outofplane} interpreted $H_c^{\mathrm{AFM}}$ as the saturation field for out-of-plane magnetization. However, the characteristic features in our magnetocaloric measurements indicate the existence of another phase transition beyond this field. Considering a single AFM-to-saturation transition scenario, below $T_\mathrm{N}$, the MCE should show a temperature minimum at the saturation field followed by rapid increase; while above $T_\mathrm{N}$, isentropic curves should rise monotonically (see black dashed curves in Fig.~\ref{Temp}(a) inset). However, our experiments reveal peculiar downward features in the isentropic curves for $H > 50$~T or T $>$ T$_{\rm N}$
as shown in Fig.~\ref{Temp}(a), which clearly deviates from the expected behavior of a single transition field scenario. As indicated by the red curves in the inset of Fig.~\ref{Temp}(a), our result suggests an alternative scenario: the emergence of an intermediate phase above $H_c^{\rm AFM}$, and another critical field at a higher field.
We find the similiar results are also reported in Ref.~\cite{Zhang2024outofplane} up to 60~T. Considering the dome-like structure of the isentropic curves are clearly observed in our experiments (as illustrated in the inset of Fig.~\ref{Temp}(a)), we only measure the MCE data up to 55~T.

To probe the upper critical field, we performed magnetization measurements up to 100~T under $c$-axis fields via the induction method. We present the magnetization process [Fig.~\ref{Temp}(b)] and its derivative d$M$/d$H$ [Figs.~\ref{Temp}(c)] with red solid curves. The absolute values are calibrated by the magnetization results up to 60~T from Ref.~\cite{Zhang2023}. The out-of-plane magnetization process exhibits substantial differences compared to the in-plane process [see inset of Fig.~\ref{Temp}(b)]. Under $H \parallel c$, four anomalies were identified at 7~T, 16~T, 37~T, and 82~T, which correspond to the peaks and shoulders in the d$M$/d$H$ curve. The first three anomalies are associated with $H_c^1$, $H_c^2$, and $H_c^{\mathrm{AFM}}$, which have also been observed in the MCE measurements. The slight numerical differences can be attributed to low-field errors from different measurement methods, as well as variations in environment temperatures. Notably,  we discover a previously unreported anomaly at 82 T ($H_c^{\rm S}$). Through replicate measurements at $\theta \simeq 0^\circ$ (Fig.~\B{S1}, Supplemental Material \B{A}~[\B{66}]), we confirm this feature is intrinsic and not an experimental artifact.

By analyzing the magnetic moment above $H_c^{\rm S}$, we conclude that this anomaly corresponds to the critical field for the saturated state. The absolute magnetic moment ($M_c^{\rm sat}$) is approximately 1.6~$\mu_B$ for the out-of-plane field direction. The in-plane saturated magnetic moment ($M_{ab}^{\rm sat}$) is reported to be 3~$\mu_B$, as shown in the inset of Fig.~\ref{Temp}(b). Considering the different in-plane and out-of-plane $g$-factors ($g_{ab}=4.13$ and $g_c=2.3$)~\cite{lin2021nc}, we find that $M_c^{\rm sat}/M_{ab}^{\rm sat} \simeq g_c/g_{ab}$~\footnote{While our results satisfy this consistency criterion, the previously reported 40 T saturation field for Na$_2$Co$_2$TeO$_6$ in Ref.~\cite{Zhang2024outofplane} fails this validation test.}, which identifies the saturation critical field $H_c^{\rm S}$ of Na$_2$Co$_2$TeO$_6$ under out-of-plane field. Combined with the critical field $H_c^{\rm AFM}$, which suppresses the AFM order, we reveal the emergence of an intermediate-field phase.

\begin{figure}[t]
\begin{center}
\includegraphics[width = 1.0\linewidth]{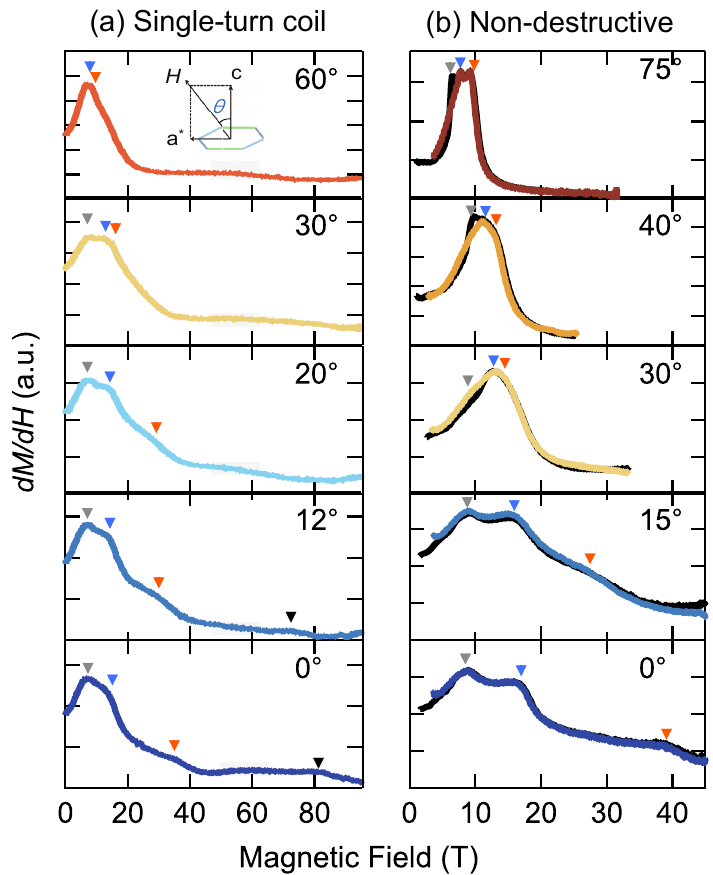}
\caption{Measured d$M$/d$H$ data versus magnetic fields under various $\theta$ angles at 4.2~K. (a) d$M$/d$H$ curves measured by the single-turn coil; (b) d$M$/d$H$ measured under the non-destructive magnetic fields. The colorful and black curves represent the down- and up-sweep results, respectively. The gray, blue, red, and black arrows correspond to the transition fields at $H_c^1$, $H_c^2$, $H_c^{\rm AFM}$, and $H_c^{\rm S}$, as in Figs.~\ref{Temp}(b) and (c). The inset illustrates the angles between the magnetic field and $c$ axis. All magnetization derivative ($dM/dH$)  panels are plotted with y-axes beginning at zero. The field-independent hump structure marked with light gray windows may caused by the experimental errors as 
shown in Fig.~\B{S1}~[\B{66}].
}
\label{Comparison}
\end{center}
\end{figure}

\begin{figure}[t]
\begin{center}
\includegraphics[width = 1\linewidth]{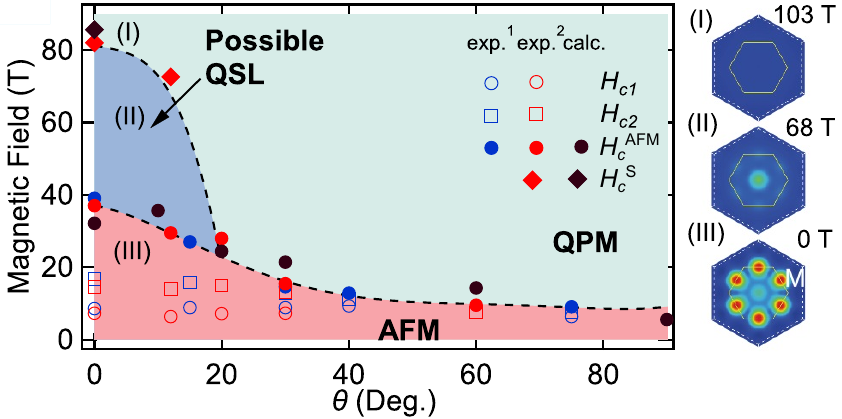}
\caption{Field-angle phase diagram determined by the low-temperature experiments and model calculations. The experimental critical fields are obtained by the magnetization measurements in Fig.~\ref{Comparison}. ``exp.$^1$'' and ``exp.$^2$'' represent the single-turn coil and non-destructive field data. We use open circle, open square, solid circle, and solid diamond to mark the transitions $H_c^1$, $H_c^2$, $H_c^{\rm AFM}$, and $H_c^{\rm S}$. The results from DMRG calculations are also indicated by solid circles and diamonds. Three phases, i.e., AFM, possible QSL, and quantum paramagnetic (QPM) states are separated by the dashed curves. Panels (I-III) in the right side show the static spin structure factors under three field strength at $\theta=0^{\circ}$.
}
\label{phase_diagram}
\end{center}
\end{figure}

In order to investigate the magnetic anisotropy, we further conducted measurements of the magnetization process along different field directions using the single-turn coil field generator. The results are shown in Fig.~\ref{Comparison}(a), where the field angle $\theta$ is tilted within the a$^{*}$-$c$ plane. To verify measurement reproducibility, we conducted complementary non-destructive field experiments up to 50~T [Fig.~\ref{Comparison}(b)], which consistently reproduced the magnetization behavior observed in destructive measurements. From the magnetization data, we find $H_c^1$ and $H_c^2$ are almost $\theta$-independent, indicating these phase transitions originate from 3D effects. Here, we focus on the two critical fields related to the intermediate phase, i.e. $H_c^{\rm AFM}$ and $H_c^{\rm S}$. $H_c^{\rm AFM}$ exhibits significantly $\theta$ dependence. At $\theta \simeq 0^\circ$, $H_c^{\rm AFM}$ is $\sim$37~T at 4.2~K, in agreement with the non-destructive result, 38~T. As $\theta$ gradually increase to 75$^{\circ}$, $H_c^{\rm AFM}$ slowly decreased to $\sim$10~T, reflecting strong magnetic anisotropy. Furthermore, we find $H_c^{\rm S}$ show even stronger magnetic anisotropy than $H_c^{\rm AFM}$ --- $H_c^{\rm S}$ becomes hardly discernible above 12$^\circ$. In Fig.~\ref{phase_diagram}, we summarize the results of magnetization measurements along different field directions in a $\theta$-$H$ phase diagram.

Here, we discuss nature of the newly observed intermediate phase between $H_c^{\rm AFM}$ and $H_c^{\rm S}$. As the ground state of AFM order has been suppressed by external fields, there could be two possibilities: a field-induced ordered phase, or an intermediate disordered phase. In the former case, the MCE curves should cross the phase boundary of the intermediate phase at a finite temperature, and feature the minima around the phase boundary~\cite{Xiang2024}, which is not supported by our MCE data in Fig.~\ref{Temp} (a). Thus the intermediate phase should more likely to be a disordered phase.

It is noteworthy that the intermediate phase is very robust against magnetic field, which extends across a wide range of external field, i.e., from 37~T to 82~T. 
The model calculations find strong quantum fluctuations in this intermediate disordered phase (see Fig.~\ref{phase_diagram}), which supports that it is a QSL phase~\cite{Zhou2017quantum}. The MCE behavior around $H_c^{\rm AFM}$ (see Fig.~\ref{Temp} (a)) also resembles the previous simulation results for $\alpha$-RuCl$_3$ model~\cite{Han2021,Han2023,Yu2023}. A dip feature observed in isentropes is found to be relatively weak near the transition point from AFM to the intermediate quantum disordered phase~\cite{Han2021}, suggesting a small entropy change, possibly ascribed to the relatively large Kitaev interaction (about 25~meV)~\cite{Han2021,Han2023,Yu2023}. Consequently, the entropy differences between different phases are rather limited in the relevant low-temperature regime. We find this phase only emerges at $\theta$ less than 12$^\circ$, which is also very similar to the possible intermediate QSL state previously reported in $\alpha$-RuCl$_3$ experiments~\cite{Zhou2023possible}. 

The establishment of a microscopic spin model is essential for elucidating the phases and phase transitions induced by an external field. Here we propose a set of parameters based on the $K$-$\Gamma$-$\Gamma^{\prime}$-$J$ spin model, i.e., $H=\sum_{\langle i,j \rangle_{\gamma}} [K S_i^{\gamma}S_j^{\gamma} + J\,\textbf{S}_i\cdot \textbf{S}_j 
+ \Gamma(S_i^{\alpha}S_j^{\beta} + S_i^{\beta}S_j^{\alpha}) + \Gamma'(S_i^{\gamma}S_j^{\alpha} 
+ S_i^{\gamma}S_j^{\beta} + S_i^{\alpha}S_j^{\gamma}+S_i^{\beta}S_j^{\gamma})]$,  
with $K = 19$~meV, $J = -0.9\left|K\right|$, $\Gamma = -0.65\left|K\right|$, 
and $\Gamma^{\prime}= 0.36\left|K\right|$,
that can capture the magnetization process of Na$_2$Co$_2$TeO$_6$, as revealed by DMRG~\cite{PhysRevLett.69.2863} results in Fig.~\ref{phase_diagram}. Our comprehensive analysis
on the calculated magnetization results of previously proposed models (see Supplemental Material \B{B}~[\B{66}]) reveals that only the Kitaev-dominant spin models can account for the emergent intermediate phase.

With the proposed model, we further calculate the spin structure factors $S({\textbf q})$ under various out-of-plane fields (see Fig.~\ref{phase_diagram} and Supplemental Materials~\B{B}~[\B{66}]). At zero and low fields, the structure factor $S({\textbf q})$ exhibits pronounced peaks at the M points of the Brillouin zone, confirming the presence of AFM order. This M-point ordering is consistent with previous neutron scattering observations~\cite{Lefran2016zigzag,Yao2022PRL}. In the intermediate-field regime, the M-point intensity is significantly suppressed, which signifies the presence of quantum spin disorder phase observed in our high-field experiments. 

In summary, we have constructed the field-angle phase diagram of Na$_2$Co$_2$TeO$_6$ through magnetization and MCE measurements at high magnetic field. At low temperatures, Na$_2$Co$_2$TeO$_6$ enters an AFM phase, showing strong magnetic anisotropy. After the AFM order is suppressed by an external field, the system enters the PM phase for $\theta \gtrsim 20^{\circ}$. Notably, at small angles $\theta$, a robust intermediate quantum disordered phase emerges between critical fields $H_c^{\rm AFM}$ and $H_c^{\rm S}$ (observed at $\theta \simeq 0^\circ$ and $12^\circ$). The upper field boundary ($H_c^4 \simeq 82$~T) demonstrates the robustness of the intermediate phase against fields, suggesting strong quantum fluctuations and potential connections to the QSL state predicted in the pure AFM Kitaev model under out-of-plane fields. 

Combining experimental data with DMRG calculations, we propose a microscopic spin model with a large AFM Kitaev interaction, which supports Na$_2$Co$_2$TeO$_6$ as a ``dual" Kitaev material to $\alpha$-RuCl$_3$ as the model parameters of the two compounds can be approximately transformed into each other through a unitary transformation (see Supplemental Materials~\B{B}~[\B{66}]). While $\alpha$-RuCl$_3$ remains the prototypical Kitaev material with dominant ferromagnetic Kitaev interactions, \ncto exhibits strikingly similar phase diagram under magnetic fields. Compared to $\alpha$-RuCl$_3$, Na$_2$Co$_2$TeO$_6$ also exhibits a double-peak feature in specific heat at zero field, with the low-$T$ peak at T$_{\rm N}$ and a broader one near 100~K~\cite{Yang2022,Yao2020PRB}. Both materials show M-point peaks in neutron scattering~\cite{Bera2017,Yao2022PRL} and display strong magnetic anisotropy~\cite{Yang2022,Xiao2021magnetic,Zhang2023}. Crucially, our MCE measurements provide evidence for quantum spin disorder in the intermediate phase of \ncto, a feature not previously experimentally established in $\alpha$-RuCl$_3$ due to technical challenges. As further high-field experiments can be conducted on \ncto, such as specific heat measurements, which are expected to exhibit power-law scaling in the intermediate-field phase~\cite{li2025kitaevderivedgaplessspinliquid}, our results highlight the unique value of \ncto as a ideal platform for exploring Kitaev physics under high fields.

% ===== Acknowledgements ==== %
\textbf{Acknowledgements} ---
X.-G.Z. thank Jian Yan, Fengfeng Song and Weiliang Yao for fruitful discussions. X.-G.Z. was supported by JSPS KAKENHI, Grant-in-Aid for Scientific Research (No. JP22H00104). X.-G.Z. and Y.H.M. was funded by JSPS KAKENHI, Grant-in-Aid for Transformative Research Areas (A) Nos.23H04859 and 23H04860, Grant-in-Aid for Scientific Research (B) No. 23H01117, and
Grant-in-Aid for Challenging Research (Pioneering) No.20K20521. 
H.L. and W.L. were supported by the National Natural Science Foundation of China (Grant Nos.~12222412 and 12447101 (W.L.), 12404177 (H.L.)), CAS Project for Young Scientists in Basic Research (Grant No. YSBR-057 (W.L.)), and the Talent Fund of Beijing Jiaotong University (Grant No. 2025JBRC003) (H.L.). Work at CHMFL(Hefei) was supported by Anhui 318
Provincial Major S\&T Project (s202305a12020005). Work at SNU was supported by the Leading Researcher Program of the National Research Foundation of Korea (Grant No. 2020R1A3B2079375).

\sloppy
%\bibliography{kitaevRef}
%apsrev4-2.bst 2019-01-14 (MD) hand-edited version of apsrev4-1.bst
%Control: key (0)
%Control: author (8) initials jnrlst
%Control: editor formatted (1) identically to author
%Control: production of article title (0) allowed
%Control: page (0) single
%Control: year (1) truncated
%Control: production of eprint (0) enabled
%

\end{document}